\documentclass[11pt,preprint]{aastex}

\newcommand{\lsun}{\,L_{\odot}}

\begin{document}

\title{Infrared Properties of Radio-Selected Submillimeter Galaxies in
the Spitzer First Look Survey Verification Field}

\author{D.\ T.\ Frayer\altaffilmark{1},
S.\ C.\ Chapman\altaffilmark{2},
L. Yan\altaffilmark{1},
L.\ Armus\altaffilmark{1},
G.\ Helou\altaffilmark{1},
D. Fadda\altaffilmark{1},
R. Morganti\altaffilmark{3}, 
M.\ A.\ Garrett\altaffilmark{4},
P. Appleton\altaffilmark{1},
P. Choi\altaffilmark{1},
F. Fang\altaffilmark{1},
I. Heinrichsen\altaffilmark{1},
M. Im\altaffilmark{5},
M. Lacy\altaffilmark{1},
F. Marleau\altaffilmark{1},
F.\ J.\ Masci\altaffilmark{1},
D.\ L.\ Shupe\altaffilmark{1},
B.\ T.\ Soifer\altaffilmark{1},
G.\ K.\ Squires\altaffilmark{1},
L.\ J.\ Storrie-Lombardi\altaffilmark{1},
J.\ A.\ Surace\altaffilmark{1},
H.\ I.\ Teplitz\altaffilmark{1},
G. Wilson\altaffilmark{1}}

\altaffiltext{1}{Spitzer Science Center, California Institute of Technology
220--06, Pasadena, CA  91125, USA} 
\altaffiltext{2}{California Institute of Technology
320--47, Pasadena, CA  91125, USA} 
\altaffiltext{3}{ASTRON, Dwingeloo, The Netherlands} 
\altaffiltext{4}{JIVE, Dwingeloo, The Netherlands} 
\altaffiltext{5}{Astronomy Program, SEES, Seoul National University, Seoul, Korea} 

\begin{abstract}

We report on submillimeter and infrared observations of 28
radio-selected galaxies in the {\it Spitzer} First Look Survey
Verification field (FLSV).  All of the radio-selected galaxies that show
evidence for emission at 850$\mu$m with SCUBA have {\it Spitzer}
counterparts at 24$\mu$m, while only half of the radio-selected galaxies
without 850$\mu$m emission have detectable counterparts at 24$\mu$m.
The data show a wide range of infrared colors (S70/S24\,$<5$--30,
S8/S3.6\,$<0.3$--4), indicative of a mixture of infrared-warm AGN and
cooler starburst dominated sources.  The galaxies showing 850$\mu$m
emission have {\it Spitzer} flux densities and flux density ratios
consistent with the range of values expected for high-redshift
($z=1$--4) ultraluminous infrared galaxies.

\end{abstract}


\section{Introduction}

The {\it Spitzer Space Telescope} provides us with the exciting
opportunity to study the high-redshift universe at mid\&far-infrared
wavelengths.  The {\it IRAS} mission first uncovered the presence of
infrared luminous galaxies in the local universe (Neugebauer et
al. 1984), and the sub-mm/mm surveys with SCUBA and MAMBO have
highlighted the importance of ultraluminous infrared galaxies (ULIRGs,
$>10^{12}\lsun$) at high redshift (e.g., Smail, Ivison, \& Blain 1997;
Hughes et al.\ 1998; Bertoldi et al. 2000).  Studies of high-redshift
ULIRGs are important for our general understanding of galaxy evolution
since they are responsible for a significant fraction of the total
energy generated by all galaxies over the history of the universe (e.g.,
Blain et al. 2002).

The recent spectroscopic studies of the sub-mm galaxy (SMG) population
show that the redshift distribution peaks at $z\sim 2$--3 (Chapman et
al.  2003b, 2004), and that the population is comprised of starbursts,
AGN, and composite AGN+starburst systems (Ivison et al. 1998, 2000;
Frayer et al. 2003; Knudsen, van der Werf, \& Jaffe 2003).  Even though
many SMGs show the presence of AGN, the molecular CO line emission
(Frayer et al. 1998, 1999; Neri et al. 2003) and X-ray data (Alexander
et al. 2003) are consistent with the majority of the infrared emission
for the population arising from star formation.  The Multiband Imaging
Photometer for {\it Spitzer} (MIPS, Rieke et al. 2004) allows us to
directly measure the infrared colors and constrain the fraction of
infrared-warm AGN dominated versus infrared-cool starburst dominated
SMGs.


\section{Observations}

Before the launch of {\it Spitzer}, we identified potential SMGs in the
First Look Survey Verification field (FLSV) by selecting radio sources
with faint optical counterparts, following previous successful selection
techniques (e.g., Cowie, Barger, \& Kneib 2002; Chapman et al. 2003a).
We used deep Westerbork 1.4\,GHz radio data (rms$=9\mu$Jy, Morganti et
al. 2004) and deep optical NOAO $R$-band data ($3\sigma=26.4$\,mag,
Fadda et al. 2004a) to derive a list of candidate sources for follow-up
observations with SCUBA.  In the Spring of 2003, we observed 28 galaxies
at the James Clerk Maxwell Telescope (JCMT) using the SCUBA
two-bolometer photometry mode, achieving rms levels of 2--3\,mJy at
850$\mu$m.

The {\it Spitzer} observations were taken as part of the Extragalactic
First Look Survey (FLS)\footnote{{\it Spitzer} program ID = 26,
http://ssc.spitzer.caltech.edu/fls/extragal/spitzer.html.}.  The 28
galaxies in the sample are located within the 0.25 degree$^{2}$ of the
FLS verification field (FLSV) and were observed with both the InfraRed
Array Camera (IRAC, Fazio et al. 2004) and MIPS.  The data presented
here have effective integration times of 480\,s, 336\,s, 168\,s, and
34\,s for the IRAC bands and the MIPS 24, 70, and 160$\mu$m arrays,
respectively.  The data were reduced using the standard SSC pipeline and
were coadded and corrected offline as needed.  The details of the data
reduction will be described in future FLS data papers (IRAC: Lacy et
al. 2004; MIPS-24$\mu$m: Fadda et al. 2004b; MIPS-70,160$\mu$m: Frayer
et al. 2004b).

\section{Results and Discussion}

\subsection{Source Identification}

We observed 28 Westerbork radio sources with SCUBA and detected 7 SMGs
at S/N$>3$.  Fourteen sources were not detected at 850$\mu$m, and the
remaining 7 sources have marginal results, showing positive signals of
1.5--3$\sigma$ (Column\,[4], Table~1). The Westerbork radio data have a
resolution of $14^{\prime\prime}\times11^{\prime\prime}$ which is well
matched to the SCUBA data, but is not sufficient to obtain reliable
counterparts in general for the optical and {\it Spitzer} data.  We used
the higher resolution ($5\farcs0$) VLA (1.4\,GHz) data of the field
(Condon et al. 2003) to obtain more accurate radio positions.  The
resolution of the VLA data is well matched to the MIPS-24 resolution of
$6^{\prime\prime}$.  Although the VLA data have lower S/N (rms$\sim
20\,\mu$Jy), the Westerbork sources were typically detected at $\ga
3\sigma$ in the VLA image, providing radio positional errors of about
1--2$\arcsec$.  The offsets between radio and MIPS-24 positions are
typically less than 2$\arcsec$, consistent within the positional
uncertainties of the radio data and the 1.5--2$\arcsec$ positional
uncertainties of the MIPS-24 data set.  Based on the 24$\mu$m counts in
the FLS field (Marleau et al. 2004), the probability of a chance
coincidence within 2$\arcsec$ is only about 1\% for 24$\mu$m sources
brighter than 100$\mu$Jy.  Hence, confusion at 24$\mu$m is not a
significant issue for this study.  After the identification of the
MIPS-24 and R-band counterparts based on the radio positions, the
corresponding MIPS-24 and optical positions were used to identify the
appropriate IRAC counterparts.

Four of the Westerbork sources have multiple radio components at the
resolution of the VLA data (1, 45, 79, 128).  Source 1 is the brightest
{\it Spitzer} source and is comprised of five components, all detected
in the MIPS-24, R-, and the IRAC-bands.  The details of the source
identification are provided in Table~1.  Figure~1 shows the MIPS-24 and
IRAC-3.6 images for the 7 SMGs detected at 850$\mu$m with S/N$>3$.

All 14 sources showing 850$\mu$m emission above the $1.5\sigma$ level
have 24$\mu$m counterparts, although source 208 is only detected at
about the 2.5--$3\sigma$ level at 24$\mu$m.  The detection of 24$\mu$m
counterparts associated even with the marginal 850$\mu$m sources gives
credence to the reliability of the $>3\sigma$ SMG sources.  In
comparison for the 14 radio sources without SCUBA detections, only half
have 24$\mu$m counterparts.  The detection of 24$\mu$m counterparts for
this sample of radio-selected SMGs is consistent with the high fraction
of 24$\mu$m counterparts found for the MAMBO and SCUBA sources in the
Lockman Hole (Ivison et al. 2004; Egami et al. 2004).

Seven sources in the full sample do not have optical counterparts down
to R$=26.4$\,mag, and only three are not detected by IRAC.  The only
source without any optical or IRAC counterpart is source\,108 which is
the brightest radio galaxy in the sample (likely a radio-loud AGN).  Two
sources (199\,[$z=1.06$] \& 150\,[$z=0.84$]) in the sample have
spectroscopic redshifts from the ongoing Keck DEIMOS redshift survey
(Choi et al. 2004).  The low redshift of $z=1.06$ for the 850$\mu$m
source\,199 implies a cool dust temperature of 20\,K assuming a
temperature dependent sub-mm to radio redshift relationship (Blain
1999).

\subsection{Infrared Properties}

The MIPS 24$\mu$m flux densities for the galaxies detected at 850$\mu$m
are consistent with expectations, assuming standard ULIRG spectral
energy distributions (SEDs) at the typical redshifts of $z\sim 2$--3
found for the SMG population (Chapman et al. 2003b; 2004).  Figure~2
shows a range of S24/S(1.4GHz) flux density ratios for the population
that may reflect a wide distribution of infrared colors.  All of the
galaxies showing 850$\mu$m emission lie within the range of values
expected for local ULIRGs redshifted to $z=1$--4.  Galaxies without
850$\mu$m emission show a wider range of S24/S(1.4GHz) flux density
ratios.  About half of the radio sources without detectable 850$\mu$m
emission have properties consistent with ULIRGs, while the other half
show excess radio emission compared to their 24$\mu$m emission,
consistent with different degrees of radio-loudness.

MIPS 70$\mu$m observations can constrain the infrared colors of the
SMGs.  Infrared-warm, AGN dominated sources are expected to have flux
density ratios of S70/S24\,$\sim 5$, while starburst dominated,
infrared-cool sources are expected to have ratios of S70/S24\,$\ga 10$.
We only detect two sources at 70$\mu$m, both of which are cool in the
infrared (source\,1: S70/S24\,$=11$ and source\,47: S70/S24\,$=26$).  To
derive an estimated average S70/S24 ratio for the SMG population, we
coadded the 70$\mu$m data at the radio positions for the sources showing
850$\mu$m emission.  We find an upper-limit of S70\,$<1.2$\,mJy
($2\sigma$).  This corresponds to an infrared color of S70/S24\,$<5$,
leaving out the two sources with 70$\mu$m detections.  Including the two
70$\mu$m detections, the average ratio for the SMG population is
S70/S24\,$<7$, which is slightly lower than expected if the population is
dominated by star-formation.  Taken at face-value, these results could
suggest that many SMGs are infrared-warm AGN, contrary to previous
conclusions that the population is dominated by star-formation.
However, the lower than expected S70/S24 ratios may arise from strong
7.7$\mu$m PAH emission redshifted into the 24$\mu$m band, given that the
median redshift for the SMG population is $z\sim2.4$ (Chapman et
al. 2003b; 2004).
 
The SEDs in the IRAC bands can be used to estimate redshifts from the
rest-frame 1.6$\mu$m peak expected for star-forming systems (e.g., Egami
et al. 2004) and to help identify AGN showing hot dust implied by their
red IRAC colors (e.g., Ivison et al. 2004).  For sources showing a
significant bump in the IRAC bands, the implied redshifts are generally
consistent (Column [11]\&[12], Table 1) with the redshift estimates
derived from the sub-mm to radio spectral index (Carilli \& Yun 1999;
Blain 1999).  The discrepant photometric redshifts for sources\,48\&119
may indicate cool dust temperatures ($T{d}<40$\,K) for these SMGs.

Of the sources with 850$\mu$m emission and IRAC detections, only
source\,47 shows a strong increase in flux density as a function of
wavelength across the IRAC bands consistent with an AGN.  However,
source\,47 also has the highest S70/S24 ratio in the sample
(S70/S24=26), suggesting that it is likely an infrared-cool starburst
that could be at $z\ga4$ and/or be highly reddened due to dust
extinction.  Hence, IRAC colors are not always conclusive for
determining the properties at longer infrared wavelengths.

\subsection{Extremely Red Objects}

A significant fraction of SMGs is comprised of extremely red objects
(EROs) (Smail et al. 1999; Frayer et al. 2004a).  The ERO definition of
$R-K>6$ (Thompson et al. 1999) corresponds to a flux density ratio of
S3.6/S(R)$\geq 85$.  Using this criterion, we identify 8 EROs (Fig. 3).
The 30\% (8/27) fraction of EROs in this sample is larger than the 8\%
found for the total micro-Jy radio population (Smail et al. 2002),
presumably since the SCUBA targets in this sample were generally
selected on the basis of being faint and red in the optical bands.
Figure~3 shows a wide range of IRAC colors for the SMG population, and
no correlation is observed between the IRAC colors and the $R-[3.6]$
colors.  Another interesting result is that most of the radio-selected
EROs without detectable 850$\mu$m emission (6/7) are bright at 24$\mu$m,
indicating that these sources are infrared-bright galaxies (likely
high-redshift ULIRGs) below the current SCUBA detection limits.

\section{Conclusions}

All of the radio-selected SMGs in the sample have {\it Spitzer}
counterparts, showing a wide range of infrared colors consistent with
ULIRGs at $z=1$--4.  The combination of accurate radio positions and
24$\mu$m detections is a powerful tool for the identification of SMGs.
More sensitive observations are required in the MIPS-70 and MIPS-160
bands to measure the infrared colors of the SMG population.  In
addition, observations with the {\it Spitzer} InfraRed Spectrograph are
needed to determine the level at which PAH emission contributes to the
MIPS-24 flux densities and to help determine the AGN fraction of the
population.  In general, {\it Spitzer} selected ULIRGs will be biased
toward AGN, except within specific redshift bins associated with the PAH
features which may be biased toward starbursts.

We thank the staff at the JCMT and the {\it Spitzer} Science Center for
their support of these observations.  This work is based in part on
observations made with the {\it Spitzer Space Telescope}, which is
operated by the Jet Propulsion Laboratory, California Institute of
Technology under NASA contract 1407.

\begin{figure}
\plotone{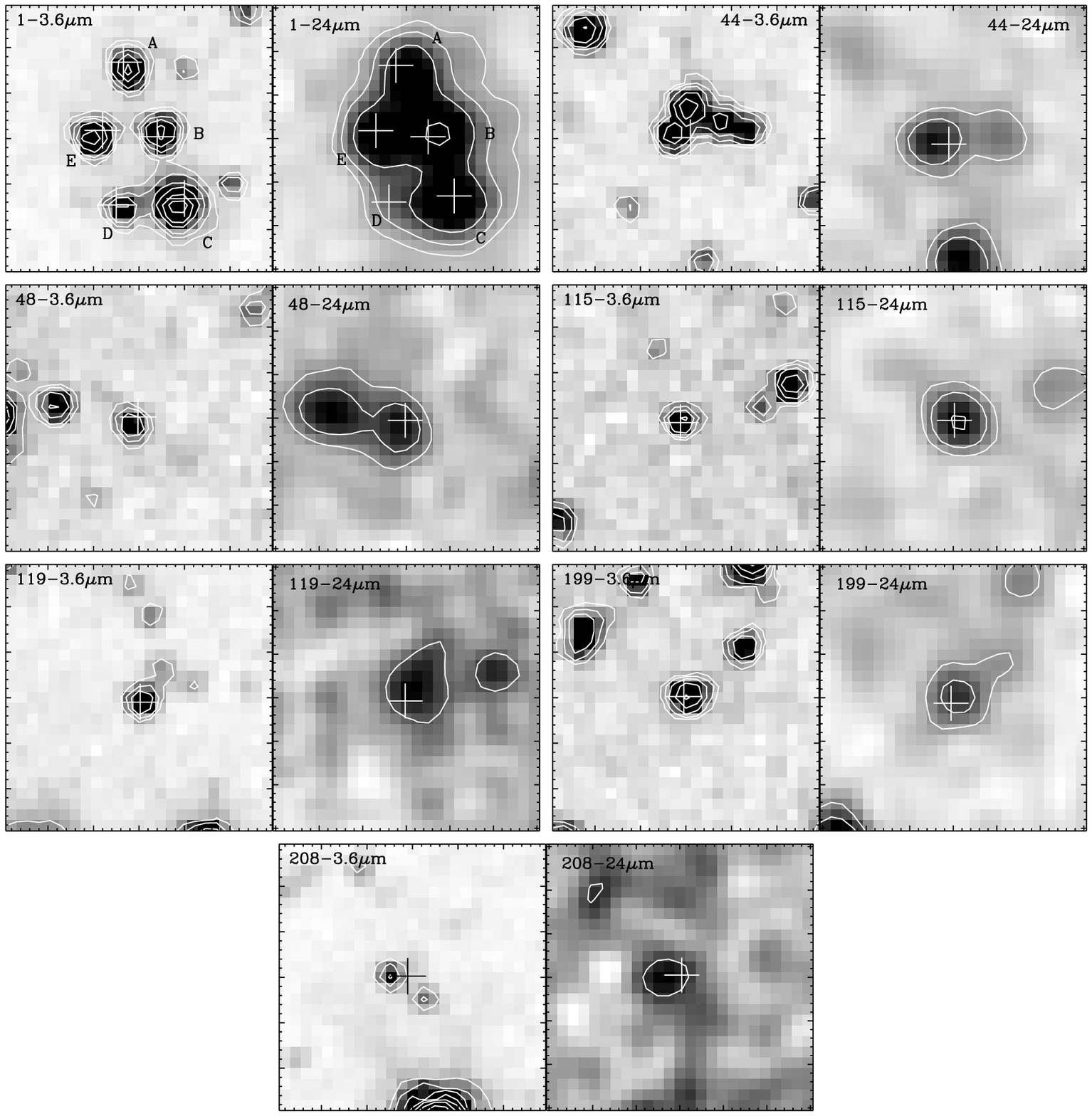}
\caption{IRAC-3.6$\mu$m and MIPS-24$\mu$m images of the seven SMGs
detected at 850$\mu$m with S/N$>3$.  Each panel is approximately
30$\arcsec$ in size (north is up and east is left) with tickmarks every
1$\arcsec$. The grey-scale is plotted on a logarithmic scale, and the
contours start at $3\sigma$ and increase by factors of 2. The crosses
represent the positions of the radio counterparts.}
\end{figure}

\begin{figure}
\plotone{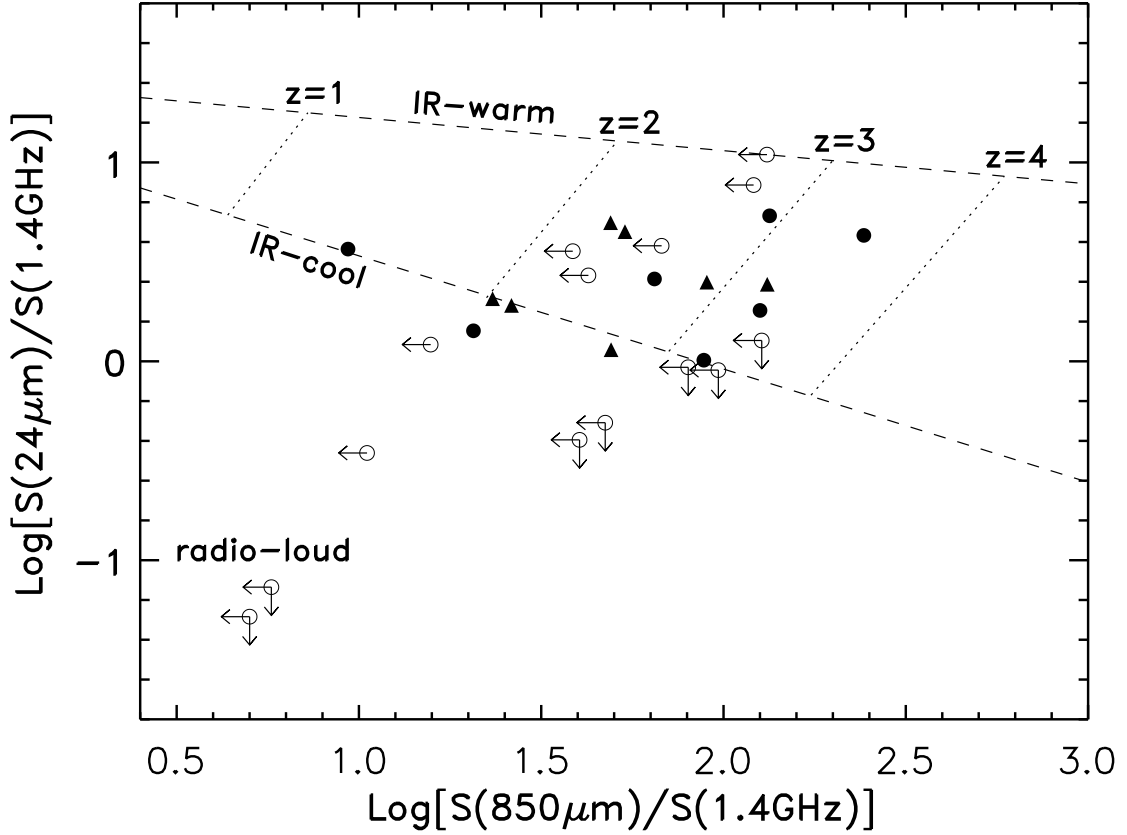}
\caption{The S24/S(1.4\,GHz) flux density ratios as a function of the
S850/S(1.4GHz) flux density ratio.  The solid circles are the $>3\sigma$
SMGs, the open circles are the radio sources without 850$\mu$m emission,
and the solid triangles are the sources with 850$\mu$m measurements of
1.5--3$\sigma$.  All limits are $2\sigma$. The dashed lines show the
expected ratios assuming power-law approximations for the infrared,
sub-mm, and radio emission.  The top dashed line represents an
infrared-warm ULIRG, while the lower dashed line represents an
infrared-cool ULIRG.  Redshift estimates are shown by the dotted lines.
The observed ratios are consistent with the SEDs of local ULIRGs at the
expected redshifts of $z\sim1$--4 for the SMG population.  Strong
radio galaxies which do not obey the far-infrared to radio correlation
are located in the lower-left.}
\end{figure}

\begin{figure}
\plotone{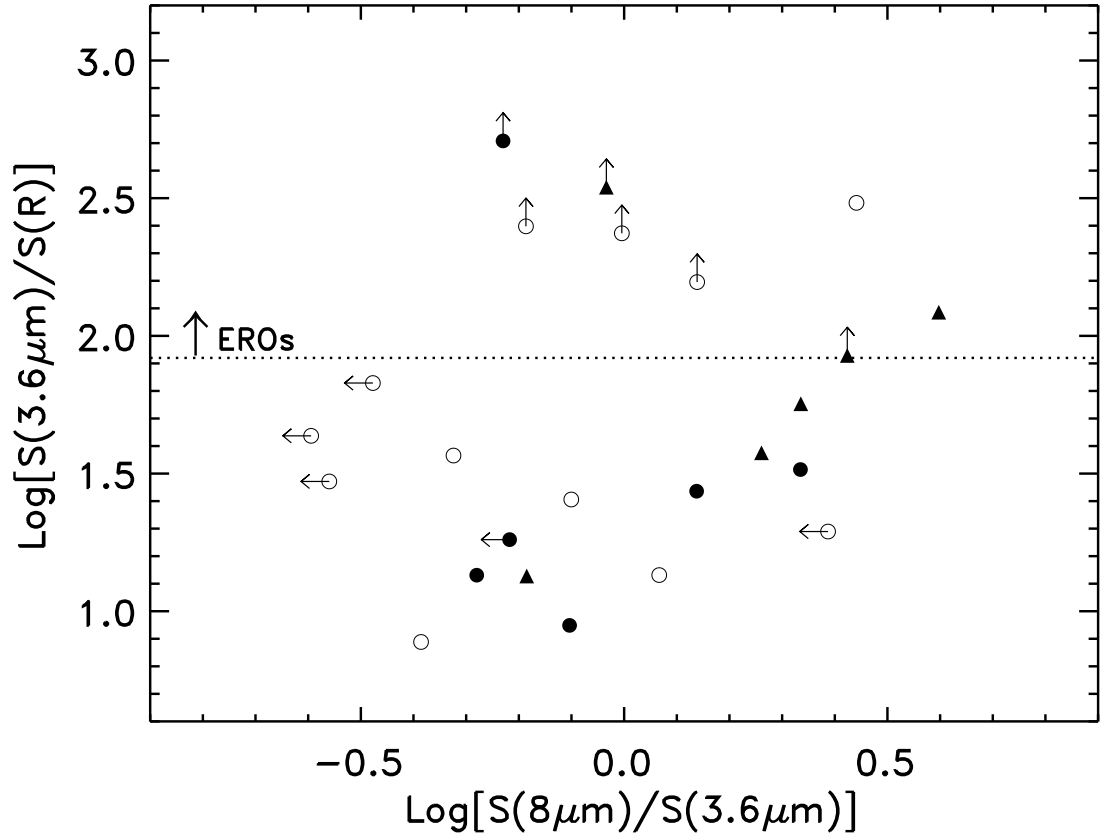}
\caption{The S3.6/S(R-band) flux density ratios as a function of IRAC
color given by the S8/S3.6 flux density ratio for sources with
detections at 3.6$\mu$m.  Symbols are the same as Figure~1. EROs are
located above the dotted line, corresponding to $R-[3.6] > 7.5$ (or
$R-K>6$).  The data show a wide range of IRAC colors.}
\end{figure}

\clearpage

\begin{deluxetable}{ccccccccccccc}
\tabletypesize{\scriptsize}
\tablewidth{0.pt}
\setlength{\tabcolsep}{0.05in}
\rotate
\tablecaption{FLSV Radio-Selected Candidate Submillimeter Galaxies}
\tablehead{
\colhead{(1)} &              
\colhead{(2)}&
\colhead{(3)}&
\colhead{(4)}&
\colhead{(5)}&    
\colhead{(6)}&
\colhead{(7)}&
\colhead{(8)}&
\colhead{(9)}&
\colhead{(10)}&
\colhead{(11)}&
\colhead{(12)}&
\colhead{(13)}\\
\colhead{Source} &              
\colhead{Position}&
\colhead{S(1.4GHz)}&
\colhead{S(850$\mu$m)}&
\colhead{S24}&    
\colhead{$R$-band}&
\colhead{S3.6}&
\colhead{S4.5}&
\colhead{S5.8}&
\colhead{S8.0}&
\colhead{submm}&
\colhead{IRAC}&
\colhead{Notes}\\
\colhead{ }&
\colhead{$\alpha$(J2000)\,$\delta$(J2000)}&
\colhead{($\mu$Jy)}&
\colhead{(mJy)}&
\colhead{($\mu$Jy)}&
\colhead{(mag)}&
\colhead{($\mu$Jy)}&
\colhead{($\mu$Jy)}&
\colhead{($\mu$Jy)}&
\colhead{($\mu$Jy)}&
\colhead{$\frac{(1+z_{phot})}{(T_d/40{\rm K})}$}&
\colhead{($1+z_{phot}$)}&
\colhead{ }}

\startdata

1&17:18:12.9\,+59:39:22&$750\pm120$&$7.0\pm2.3$&$2750\pm280$&$19.68\pm0.05$& $559\pm56$&$365\pm37$&$362\pm36$&$293\pm29$         &$2.3\pm0.8$&$<2.5$&a\\
44&17:17:29.7\,+59:54:29&$74\pm9$&$9.9\pm2.9$&$399\pm40$&$23.44\pm0.10$& \nodata&\nodata&\nodata&\nodata                         &$4.4\pm0.8$&\nodata&b\\
48&17:17:33.7\,+59:53:56&$54\pm9$&$13.1\pm2.9$&$232\pm28$&$23.83\pm0.11$&  $16.5\pm1.7$&$18.0\pm1.8$&$14.9\pm3.3$&$<10$          &$5.0\pm0.8$&$3\pm0.8$&\\
115&17:17:43.4\,+59:48:03&$262\pm26$&$5.4\pm1.6$&$373\pm37$&$24.72\pm0.20$&  $13.0\pm1.3$&$13.0\pm1.3$&$<10.0$&$28.1\pm3.3$       &$2.8\pm0.8$&\nodata&\\
119&17:17:12.7\,+59:47:53&$81\pm9$&$10.2\pm2.8$&$146\pm28$&$>26.4$&    $43.3\pm4.3$&$44.0\pm4.4$&$42.9\pm4.3$&$25.5\pm3.3$      &$4.3\pm0.8$&$3\pm0.8$&c\\
199&17:17:29.4\,+59:41:13&$116\pm12$&$7.5\pm2.5$&$301\pm30$&$23.44\pm0.14$&  $35.4\pm3.5$&$39.9\pm4.0$&$49.3\pm4.9$&$48.6\pm4.9$ &$3.7\pm0.8$&\nodata&$z=1.06^{d}$\\
208&17:18:10.9\,+59:40:41&$77\pm9$&$6.8\pm1.9$&$78\pm28$&$22.62\pm0.10$& $24.5\pm2.5$&$24.0\pm2.4$&$<10$&$19.3\pm3.3$           &$3.9\pm0.8$&\nodata&\\
\tableline
45A&17:17:47.5\,+59:54:24&$80\pm9$&$7.2\pm2.5$&$200\pm28$&$24.12\pm0.14$& $26.0\pm2.6$&$27.4\pm2.7$&$36.5\pm3.7$&$47.4\pm4.7$   &$4.0\pm0.8$&\nodata&e\\
47&17:17:22.5\,+59:54:12&$100\pm10$&$4.9\pm2.4$&$498\pm50$&$25.31\pm0.22$& $28.2\pm2.8$&$45.6\pm4.6$&$68.1\pm6.8$&$111.6\pm11.2$    &$3.4\pm0.8$&\nodata&c,f\\
73&17:17:57.9\,+59:52:00&$118\pm12$&$5.8\pm2.1$&$135\pm28$&$>26.4$& $29.4\pm2.9$&$35.5\pm3.6$&$30.7\pm3.3$&$27.2\pm3.3$          &$3.4\pm0.8$&$3\pm0.8$&c\\
139&17:18:23.2\,+59:45:53&$215\pm22$&$5.0\pm2.8$&$444\pm44$&$22.62\pm0.05$& $36.9\pm3.7$&$32.5\pm3.3$&$27.9\pm3.3$&$24.1\pm3.3$  &$2.9\pm0.8$&$<2.5$&\\
145&17:17:46.2\,+59:45:17&$66\pm9$&$8.7\pm4.0$&$161\pm28$&$>26.4$& $7.2\pm1.0$&$7.7\pm1.0$&$<10$&$19.1\pm3.3$                   &$4.3\pm0.8$&\nodata&c\\
156&17:18:16.8\,+59:44:30&$82\pm9$&$4.4\pm2.5$&$368\pm37$&$25.64\pm0.37$&  $9.7\pm1.0$&$12.9\pm1.3$&$25.0\pm3.3$&$21.0\pm3.3$    &$3.5\pm0.8$&$3.5\pm0.8$&\\
191&17:17:15.5\,+59:42:02&$145\pm15$&$3.8\pm2.5$&$277\pm28$&$25.35\pm0.34$& $<3$&$<3$&$<10$&$<10$                          &$3.0\pm0.8$&\nodata&\\
\tableline
75&17:18:01.7\,+59:51:47&$38\pm9$&$<7.5$&$416\pm42$&$>26.4$&  $20.0\pm2.0$&$24.6\pm2.5$&$27.2\pm3.3$&$19.8\pm3.3$                &$<4.8$&$3.5\pm0.8$&c\\
79A&17:17:22.8\,+59:51:30&$114\pm11$&$<6.6$&$408\pm41$&$23.30\pm0.17$&  $54.2\pm5.4$&$38.9\pm3.9$&$36.6\pm3.7$&$25.7\pm3.3$     &$<3.6$&$<2.5$&g\\
85&17:18:12.3\,+59:50:56&$44\pm9$&$<8.4$&$<84$&$21.93\pm0.05$&  $40.1\pm4.0$&$26.4\pm2.6$&$23.9\pm3.3$&$16.5\pm3.3$              &$<4.7$&$<2.5$&\\
91&17:17:43.7\,+59:50:22&$43\pm9$&$<7.8$&$331\pm33$&$>26.4$&  $13.3\pm1.3$&$17.9\pm1.8$&$15.5\pm3.3$&$18.3\pm3.3$               &$<4.7$&\nodata&c\\
99&17:17:06.4\,+59:49:25&$139\pm14$&$<8.4$&$<84$&$23.50\pm0.15$&  $36.3\pm3.6$&$25.3\pm2.5$&$10.0\pm3.3$&$<10$                  &$<3.6$&$<2.5$&h\\
108&17:17:41.2\,+59:48:36&$1080\pm130$&$<8.1$&$<84$&$>26.4$&$<3$&$<3$&$<10$&$<10$                                         &$<2.2$&\nodata&\\
109&17:17:38.5\,+59:48:32&$765\pm110$&$<6.6$&$<84$&$24.19\pm0.14$&$<3$&$<3$&$<10$&$<10$                                   &$<2.2$&\nodata&\\
128B&17:17:47.0\,+59:47:12&$62\pm9$&$<9.0$&$<84$&$25.41\pm0.26$&  $4.1\pm1.0$&$3.6\pm1.0$&$<10$&$<10$                        &$<4.4$&\nodata&i\\
155&17:17:56.0\,+59:44:32&$60\pm9$&$<7.2$&$<84$&$24.60\pm0.27$&  $30.0\pm3$&$32.7\pm3.3$&$21.7\pm3.3$&$<10$                  &$<4.3$&$3\pm0.8$&\\
136&17:17:11.6\,+59:46:21&$108\pm9$&$<6.9$&$292\pm29$&$23.89\pm0.16$&  $21.8\pm2.2$&$23.5\pm2.4$&$19.0\pm3.3$&$17.3\pm3.3$     &$<3.7$&$3\pm0.8$&\\
146&17:18:12.9\,+59:44:54&$627\pm63$&$<9.9$&$217\pm28$&$23.83\pm0.10$&  $39.3\pm3.9$&$27.4\pm2.7$&$15.7\pm3.3$&$<10$            &$<2.6$&$<2.5$&\\
150&17:17:42.4\,+59:44:56&$68\pm9$&$<6.9$&$259\pm28$&$23.37\pm0.14$&  $18.7\pm1.9$&$11.8\pm1.2$&$<10$&$21.8\pm3.3$              &$<4.1$&\nodata&$z=0.84^{d}$\\
198&17:17:21.4\,+59:41:13&$114\pm12$&$<8.1$&$<84$&$>26.4$&   $21.2\pm2.1$&$22.0\pm2.2$&$13.3\pm3.3$&$13.8\pm3.3$                  &$<3.7$&$<2.5$&c\\
211&17:18:21.3\,+59:40:27&$356\pm36$&$<8.4$&$432\pm43$&$26.16\pm0.42$&  $32.3\pm3.2$&$41.7\pm4.2$&$60.1\pm6.0$&$89.2\pm8.9$     &$<2.9$&\nodata&c\\
\enddata

\tablecomments{(1)Radio source identifier.  (2)Radio positions which are
accurate to about $\pm2^{\prime\prime}$. (3)Radio flux densities from
the Westerbork and VLA data. (4)SCUBA 850$\mu$m measurements.  Sources
1--208 have $>3\sigma$ 850$\mu$m detections.  Sources 45A--191 have
1.5--3$\sigma$ 850$\mu$m measurements, and sources 75--211 are
non-detections at 850$\mu$m.  All flux density and magnitude limits are
given as $3\sigma$. (5)MIPS-24$\mu$m measurements. (6)Optical $R$-band
magnitudes from NOAO 4m. (7)--(10)IRAC flux densities. (11)Photometric
redshift estimate of $(1+z)\times(T_{d}/40{\rm K})^{-1}$ based on the
sub-mm to radio relationship of Carilli \& Yun (1999), accounting for
the degeneracy of redshift and dust temperature in the relationship
(Blain et al. 1999). (12)Photometric redshift based on the assumed
rest-frame 1.6$\mu$m peak in the IRAC SED. (a)Measurements summed over
all five components (Fig. 1).  Source\,1 is also detected at longer
wavelengths with S70$=30\pm6$\,mJy and S160$=130\pm26$\,mJy. Sources not
detected in the 70\&160$\mu$m MIPS-bands have $3\sigma$ limits of about
$<10$\&$<50$\,mJy, respectively.  (b)Detected in IRAC-bands, but IRAC
measurements are confused with nearby bright optical source. (c)ERO
based on high S3.6/S(R) flux density ratio. (d)Spectroscopic redshift
based on Keck DEIMOS data (Choi et al. 2004). (e)Brighter radio
component 45A has 24$\mu$m counterpart and is adopted as the primary
counterpart, since 45B has no associated 24$\mu$m source. (f)Detected in
MIPS-70 with S70$=13\pm3$\,mJy.  (g)Brighter radio component 79A has a
24$\mu$m counterpart and `merger-like'' morphology with two optical
components.  (h)Source\,99 has two optical components showing a
merger-like morphology.  (i)Neither radio component of source\,128 has a
24$\mu$m counterpart; 128B is adopted as the counterpart for flux
density comparisons since there are no optical or IRAC counterparts
associated with 128A.}

\end{deluxetable}

\end{document}